\documentstyle[12pt]{article}
\begin{document}
\bibliographystyle{unsrt}
\newcommand{\bra}[1]{\left < \halfthin #1 \right |\halfthin}
\newcommand{\ket}[1]{\left | \halfthin #1 \halfthin \right >}
\newcommand{\be}{\begin{equation}}
\newcommand{\ee}{\end{equation}}

\title{\bf  On The $Q^2$ Dependence of The Spin Structure Function
\\ In The Resonance Region }

\author{Zhenping Li$^1$ and  Zhujun Li$^2$
\\
$^1$Physics Department, Carnegie-Mellon University \\
Pittsburgh, PA. 15213-3890 \\
\\
$^2$CEBAF, 12000 Jefferson Avenue \\
Newport News, VA 23606}
\maketitle

\begin{abstract}
\baselineskip=24pt
In this paper, we show what we can learn from the CEBAF experiments on
spin-structure functions, and the transition from the Drell-Hearn-Gerasimov
sum rule in the real photon limit to the spin dependent sum rules in the
deep inelastic scattering, and how the asymmetry $A_1(x,Q^2)$ approaches
the scaling limit in the resonance region.  The spin structure function
in the resonance region alone can not determine the spin-dependent
sum rule due to the kinematic restriction of the resonance
region.   The integral
$\int_0^1 \frac {A_1(x,Q^2)F_2(x,Q^2)}{2x(1+R(x,Q^2))}dx$
is estimated from $Q^2=0$ to $2.5$ GeV$^2$.  The result shows that
there is a region where both contributions from the baryon resonances
and the deep inelastic scattering are important, thus provides important
information on the high twist effects on the spin dependent
sum rule.
\end{abstract}
PACS numbers: 13.60.Hb, 11.50.Li, 12.40.Aa,14.20Gk

\newpage
\baselineskip=24pt
\subsection*{1. Introduction}
The polarized lepton nucleon scattering is determined by two
spin structure functions $g_{1,2}(x,Q^2)$, which can be
written as\cite{IOFFE}
\begin{equation}\label{21}
g_1(x,Q^2)=\frac {MK}{8\pi^2 \alpha(1+\frac {Q^2}{\nu^2})} \left [
\sigma_{1/2}(\nu,Q^2)-\sigma_{3/2}(\nu,Q^2)+\frac {2\sqrt {Q^2}}{\nu}
\sigma_{TS}(\nu,Q^2)\right ]
\end{equation}
and
\begin{equation}\label{22}
g_2(x,Q^2)=\frac {MK}{8\pi^2\alpha(1+\frac {Q^2}{\nu^2})}\left [
\frac {2\nu}{\sqrt{Q^2}}\sigma_{TS}(\nu,Q^2)-(\sigma_{1/2}(\nu,Q^2)-
\sigma_{3/2}(\nu,Q^2))\right ]
\end{equation}
where $\sigma_{1/2}(\nu,Q^2)$ and $\sigma_{3/2}(\nu,Q^2)$ are the
cross sections for photon scattering with spin of the photon parallel and
antiparallel to the spin of the longitudinally-polarized nucleon with
mass $M$, $\sigma_{TS}(\nu,Q^2)$ represents the interference between
the transverse and the longitudinal cross section, and K is the
photon flux.  The
parton model tells us that the spin dependent structure functions
$g_{1,2}(x,Q^2)$ should have simple scaling behaviors in the
large $Q^2$ limit;
\begin{equation}\label{23}
\lim_{Q^2\to \infty} g_{1,2}(x,Q^2)\to g_{1,2}(x),
\end{equation}
and the corresponding spin dependent sum rules\cite{BJ,EJ}
\begin{equation}\label{24}
\Gamma=\int_0^1 g_1(x) dx
\end{equation}
provide us an important information on the spin-flavor correlations
of the nucleon.   However, the average $Q^2$ of the polarized
lepton-nucleon scattering experiments may not be high enough
to reach the scaling limit, and  the $Q^2$ dependence of the
spin structure function becomes significant.
  Recent studies\cite{CR93,EK93,ZPLI94} have shown that
the $Q^2$ dependence of the spin structure function should be taken into
account  in order to explain
SMC\cite{SMC93} and SLAC\cite{SLAC93} measurements
on the spin structure function of the nucleon along with the EMC\cite{EMC}
data.  Theoretically, such $Q^2$ dependence that comes from the high
twist effects have been the subject of many studies\cite{BAL,JI}.
The information on the spin structure function in the low $Q^2$
region would be more interesting since the high twist effects
become stronger, this is where the spin structure function in the
resonance region plays an important role.
On the other hand, it has been shown\cite{AIL} that $Q^2$ dependence of the
spin structure function arises naturally due to the transition
from a negative Drell-Hearn-Gerasimov (DHG) sum rule\cite{DHG66}
to the positive spin-dependent sum rule in the large $Q^2$
limit.  The investigations in the explicit quark model show\cite{ZPLI93}
that the DHG sum rule is dominated by the spin structure function
in the resonance region, in particular, by the process
$\gamma_v N\to P_{33}(1232)$, which has been confirmed by the
data\cite{BURKERT}.  How this transition happens in the small
$Q^2$ region also provides an important test ground for theories,
since the explicit quark model\cite{ZPLI93} and chiral perturbation
theory\cite{BVM} have given different predictions.

The focus of this paper is to examine the $Q^2$ dependence
of the spin structure function in the resonance region.
Our investigation is based on the earlier
studies of the $Q^2$ extensions
of the DHG sum rule\cite{BURKERT,BURKERT1}, and  we further
replace the photon energy $\nu$ by variable $x$ which is analogue
to the Bjorken scaling variable in the deep inelastic scattering.
We shall show that
the resonance region alone is not sufficient to provide an accurate
estimate of the DHG sum rule due  to the kinematic restriction.
The evolution of the spin asymmetry $A_1(x,Q^2)$,
 which is defined as
\begin{equation}\label{27}
A_1(x,Q^2)=\frac {\sigma_{1/2}(x,Q^2)-\sigma_{3/2}(x,Q^2)}{\sigma_{1/2}
(x,Q^2)+\sigma_{3/2}(x,Q^2)},
\end{equation}
will also be studied in this framework.
A negative DHG sum rule at the real photon limit requires
that the spin asymmetry function $A_1(x,Q^2=0)$
 be negative at the real photon limit.  This leads to a transition
from a negative $A_1(x,Q^2)$ at the real photon limit to the
positive $A_1(x,Q^2)$ at large $Q^2$, the data\cite{BAUM}
tell us that such a transition should happen at a relative
small $Q^2$.  If this is indeed to be the case, the $A_1(x,Q^2)$
in the resonance region will show how the transition occur.
Furthermore, because the DHG sum rule is dominated by the
exclusive process $\gamma_v N\to P_{33}(1232)$, one
should expect a dramatic $Q^2$ dependence in the small $Q^2$
region.  Thus, there might be a region that
both contributions from the resonance region and deep inelastic
 scattering are important, where the $Q^2$ dependence of the structure
function is dominated by the leading high twist effects.
Our investigation shows that this is possible
for $1.0 \le Q^2 \le 2.5$ GeV$^2$. In the next section, we will
establish the connection between the DHG sum rule and the
integral of the spin structure function.
The $Q^2$ evolution of the asymmetry $A_1(x,Q^2)$ will also be shown.

Eqs. \ref{21} and \ref{22} show that the
DHG sum rule integral $\int_{\nu_{th}}^{\infty} \frac {\sigma_{1/2}
(\nu,Q^2)-\sigma_{3/2}(\nu,Q^2)}{\nu}d\nu$ used in Refs.
\cite{BURKERT,BURKERT1} is not related to the
spin dependent sum rule $\int_0^1 g_1(x,Q^2)dx$ at large $Q^2$
since the photon
flux $K$ in Eqs. \ref{21} and \ref{22} is defined as the equivalent
photon energy.  Therefore, one should estimate the integral
$I(Q^2)=\int_0^1 \frac {A_1(x,Q^2)F_2(x,Q^2)}{2x(1+R(x,Q^2))}dx$,
which is related the DHG sum rule in the real photon limit and
the spin dependent sum rules in the deep inelastic scattering
region.  Furthermore, one has to go beyond the resonance
region to give an accurate estimate of the integral $I(Q^2)$.
 This will be given
in section 3.  Finally, the conclusion will be given in
section 4.

\subsection*{\bf 2. The  Kinematics}
In order to describe the strong
$Q^2$ dependence around $Q^2=0$ of the spin structure function
due to the pion photo and electroproduction that dominate the
DHG sum rule, it is more appropriate to define the variable
$x$ as
\begin{equation}\label{1}
x=\frac {\nu_{th}}{\nu}
\end{equation}
where $\nu_{th}$ is the threshold energy for the
pion-electroproduction, which is related to the
pion mass $m_{\pi}$ and the nucleon mass $M$ by
\begin{equation}\label{2}
\nu_{th}=\frac{Q^2+2m_{\pi}M+m_{\pi}^2}{2M}.
\end{equation}
The variable $x$ defined in Eq. \ref{1} is exactly the same as
variable $\xi$ in Ref. \cite{DL}, which
gives a better description of the strong $Q^2$ dependence of the
structure function $f_2(x,Q^2)$ near threshold.  Furthermore,
 the variable $x$ naturally leads to the Bjorken scaling
variable $x_B=\frac {Q^2}{2M\nu}$ in the large $Q^2$ limit, and
makes it possible to write
the DHG integral $\int_{\nu_{th}}^{\infty}\frac {\sigma_{1/2}(\nu)-
\sigma_{3/2}(\nu)}{\nu} d\nu$ in terms of the integral
 $\int_0^1 g_1(x,Q^2)dx$ at the real photon limit;
\begin{equation}\label{5}
\int^1_0g_1(x,Q^2=0)dx =
\frac {M\nu_{th}}{8\pi^2\alpha}\int^{\infty}_{\nu_{th}}
\frac {\sigma_{1/2}(\nu,Q^2=0)-\sigma_{3/2}(\nu,Q^2=0)}{\nu} d\nu,
\end{equation}
where the photon flux $K$ in Eq. \ref{21} is the photon energy $\nu$
at the real photon limit.
The DHG sum rule gives
\begin{equation}\label{6}
\int_0^1 g_1(x,Q^2=0)dx=-\frac {\nu_{th}}{4M}\kappa^2=-\frac{2m_{\pi}M
+m_{\pi}^2}{8M^2} \kappa^2
\end{equation}
where $\kappa$ is the anomalous
magnetic moments of the nucleon. Substitute $\kappa_p=1.79$ and
$\kappa_n=-1.91$ for both protons and neutrons into Eq. \ref{6},
we have
\begin{equation}\label{7}
\int_0^1 g_1(x,Q^2=0)dx
=\left \{ \begin{array}{c} -0.123 \qquad \mbox{for protons}\\
-0.14 \qquad \mbox{for neutrons.}\end{array}\right .
\end{equation}

Replacing the photon energy $\nu$ by the variable $x$
at low $Q^2$ also shows kinematic restrictions of the
resonance region;  rewriting the photon energy as
\begin{equation}\label{3}
\nu=\frac {W^2+Q^2-M^2}{2M},
\end{equation}
where $W$ is the mass of final state when nucleon $M$ at rest
absorbs the incoming photon with energy $\nu$,
the resonance region corresponds to
the final state mass $W$ from the threshold $W=M+m_{\pi}=1.073$
 to $W=1.8$ GeV which is also the region that can be reached
in the future CEBAF experiments\cite{BEXP}.
In Fig. 1, we show the range of the
variable $x$ that is accessible to the resonance region;
at $Q^2=0$, the resonance region covers $x=0.11 \sim 1$,
while $x$ is only limited to $0.57\sim 1$ at $Q^2=2.5$
GeV$^2$.  This raises a serious question whether the resonance
region, or the information from the CEBAF experiments, alone
would be sufficient to provide an accurate estimate of the
spin dependent integral $\int_{0}^1 g_1(x,Q^2) dx$ even at moderate $Q^2$.

However, the studies of the spin structure function
in the resonance region could provide us very
useful information on transition of the spin structure function
from the real photon limit to the scaling
limit in the deep inelastic scattering.  In particular,
the data on the spin asymmetry function $A_1(x,Q^2)$ in the resonance
region will reveal how it reaches the scaling limit, if
$A_1(x,Q^2)$ has little dependence on $Q^2$ after
$Q^2\ge 0.5$ GeV$^2$.   A significant amount of the
pion and $\eta$ electroproduction data has been collected
in the resonance region\cite{FOSTER}, this enable us to extract the
transverse photo coupling amplitudes $A_{1/2}(Q^2)$ and
$A_{3/2}(Q^2)$ for the most prominent resonances in the range
$0\le Q^2 \le 3.0$ GeV$^2$.  For the other relative weak resonances,
we used the single quark transition model to calculate their photo coupling
constants. Therefore, combining the
contribution from the resonances and the background
pion production, one can obtain the spin structure function
in the resonance region, (for further detail, see \cite{BURKERT}).
In the figure 2, we show that asymmetry $A_1(x,Q^2)$ of proton
in the resonance region at $Q^2=0$, $0.5$ and $1.5$ GeV$^2$,
and the data at $Q^2=0.5$ and $1.5$ GeV$^2$\cite{BAUM} are also shown.
The asymmetry is mostly negative in the region $x\le 0.53$ in the
real  photon limit,  since it is dominated by the process
$\gamma N\to P_{33}(1232)$.  Moreover, the cancellation
between the transitions of
spin flip and orbital angular momentum flip for the resonance
$F_{15}(1688)$ and $D_{13}(1520)$\cite{FG,ZPLI92}
also makes helicity amplitudes
$A_{1/2}$ vanishes for these resonances, which leads to a negative
$A_1(x,Q^2=0)$ for the higher resonances.  As $Q^2$ increases,
the resonance $P_{33}(1232)$ is shifting to larger x,
and the magnetic spin flip transition becomes dominant,  thus
the asymmetry $A_1(x,Q^2)$ becomes positive in the small x region,
and changes the sign and approaches to unity as $x\to 1$.
This tells us that the asymmetry $A_1(x,Q^2)$ approaches to the
scaling limit in the small $x$ region first, and the range
of variable $x$ that asymmetry $A_1(x,Q^2)$ scales increases
as $Q^2$ increases.   At the scaling limit, the valence quark
model\cite{CLOSE93} predicts
\begin{equation}\label{28}
A_1(x)=\frac {19-16 R^{np}(x)}{15},
\end{equation}
where $R^{np}(x)=\frac {f_2^n(x)}{f^p_2(x)}$ is the ratio between
the structure functions of neutrons and protons, which give
a very good description of the EMC data\cite{EMC}
in the large $x$ region.  Comparing with the asymmetry $A_1(x)$ in the
scaling limit, which is also shown in Fig. 2, we find that
the $A_1(x,Q^2)$ in the resonance region does not reach the scaling
limit at $Q^2=0.5$ GeV$^2$, and it  is much closer to
the scaling limit at $Q^2=1.5$ GeV$^2$ for $x\le 0.6$.
The future experiments at CEBAF\cite{BEXP} would certainly provide the
answer in this regard.

\subsection*{\bf 3. The $Q^2$ dependence of the spin-dependent sum rule}

There are several problems  in estimate the $Q^2$ dependence
of the integral $\int_0^1 g_1(x,Q^2) dx$.
First; there is little experimental information available
on the longitudinal photon coupling $S_{1/2}(\nu,Q^2)$
in the resonance region, we can only estimate  the
integral
\begin{eqnarray}\label{25}
I_p(Q^2) & = & \int_0^1 \frac {MK}{8\pi^2\alpha (1+\frac {Q^2}{\nu^2})}
 \left (\sigma_{1/2}
(x,Q^2)-\sigma_{3/2}(x,Q^2)\right ) dx \\ \nonumber & = &
\int_0^1 \frac {A_1(x,Q^2)f_2(x,Q^2)}{2x(1+R(x,Q^2))} dx,
\end{eqnarray}
where $R(x,Q^2)=\sigma_L(x,Q^2)/\sigma_T(x,Q^2)$ is the ratio
between the longitudinal and transverse cross sections, and $K$
is defined as the equivalent photon energy.  This integral equals
$\int_0^1g_1(x,Q^2)dx$ at the real photon limit and the large
$Q^2$ limit.
Soffer and Teryaev\cite{SOFFER} pointed out that the
strong $Q^2$ dependence  of the spin structure function $g_1(x,Q^2)$
 in the small $Q^2$ region might be
due to the contamination of the function $g_2(x,Q^2)$, this means
that the quantity $\sigma_{TS}(\nu,Q^2)$ in Eq. \ref{21}
 might play an important role
in the small $Q^2$ region.  This might be related to the threshold effect
around $Q^2=0$, thus one expected it vanish very quickly
as $Q^2$ increases.  The quantitative calculation of
$\sigma_{TS}(\nu,Q^2)$ can be done in the framework of valence quark
model, and this is in progress.
Second, the kinematic restriction shows that the spin structure
function  in the resonance region
alone might not be sufficient to estimate the integral $I_p(Q^2)$,
Fig. 3 shows the quantity
$I(x_{min})=\int^1_{x_{min}}
\frac {A_1(x,Q^2)f_2(x,Q^2)}{2x(1+R(x,Q^2))} d x$ as a function
of $x_{min}$ at $Q^2=0$. The DHG sum rule
should be extracted at $x_{min}=0$, and if the function $I(x_{min})$
reaches constant in the small $x_{min}$ region, it means
that the integral $\int^{x_{min}}_0 g_1(x,Q^2)dx$ can be neglected,
thus an accurate DHG sum rule can be obtained experimentally.
Unfortunately, Fig. 3 shows that $x_{min}$ is not small enough to warrant
such extraction, and $I(x_{min})$ has not become constant
in the small $x_{min}$ region.  Therefore, there is a theoretical
uncertainty in the estimate of the DHG sum rule due to the
kinematic restrictions of the resonance region, which has not been
discussed in the literature\cite{IK73}.  The DHG could be obtained
by averaging two extreme cases; extrapolating $I(x_{min}=0)$ from
the function $I(x_{min})$ at small $x_{min}$ region assuming
a smooth behavior of the function $I(x_{min})$ near $x_{min}=0$,
or neglecting the contribution from the outside of the
resonance region,  and the difference between these two cases
represents the theoretical uncertainty.  Following this
procedure, we find
\begin{equation}\label{33}
\int^1_0 g_1^p(x,Q^2=0)dx=-0.126 \pm 0.011.
\end{equation}
This result is in good agreement with the theoretical prediction for
protons in Eq. \ref{7}, and the theoretical uncertainty is quite
significant.

Therefore, one can generally write the integral for the spin
dependent structure function $I_p(Q^2)$ as
\begin{equation}\label{8}
I_p(Q^2)=I^{res}(Q^2)+I^{nonres}(Q^2)=\int_{x_r}^1+\int_0^{x_r}
\frac {A_1(x,Q^2)f_2(x,Q^2)}{2x(1+R(x,Q^2))}dx,
\end{equation}
where $I^{res}(Q^2)$ represents the contribution from the resonance
region between $x_{r}$ and 1, and $I^{nonres}(Q^2)$ comes from the
outside resonance region.  The quantity $x_r$ in Eq. \ref{8} is
\begin{equation}\label{26}
x_r=\frac {Q^2+2m_{\pi}M+m_{\pi}^2}{W_r^2+Q^2-M^2}
\end{equation}
with $W_r=1.8$ GeV.  Thus, $x_r$ approaches to unity as $Q^2$ increases,
and consequently the contribution from the resonance region decreases.
    Our result shows that the asymmetry $A_1(x,Q^2)$
outside the resonance region may not reach the scaling limit
for $x\le x_r$
until $Q^2\approx 1$ GeV$^2$, where the asymmetry data in deep inelastic
scattering\cite{EMC} could be used.
Fig. 4 shows our estimate the $Q^2$ dependence of the integral
$I(Q^2)$.  The qualitative behavior of $I^{res}(Q^2)$
is consistent with the results in Refs. \cite{BURKERT,BURKERT1},
there is a strong $Q^2$ dependence of $I^{res}(Q^2)$ in the region
$0\le Q^2 \le 0.5$ GeV$^2$ due to the transition
$\gamma_v N\to P_{33}(1232)$.  $I^{res}(Q^2)$ becomes positive at $Q^2
\approx 0.6$ GeV$^2$ and reaches maximum at $Q^2=1.25$ GeV$^2$.
However, $I^{res}(Q^2)$ is not the whole story, one has to add
$I^{nonres}(Q^2)$ to give an accurate estimate of $I(Q^2)$.
Since the asymmetry $A_1(x,Q^2)$ in the resonance region
shows a strong $Q^2$ dependence in $Q^2\le 1$ GeV$^2$,
we only add $I^{nonres}(Q^2)$ after
$Q^2\ge 0.75$ GeV$^2$,
in which $A_1(x)$ is taken from the
EMC data\cite{EMC}, the structure function $f_2(x,Q^2)$ is
obtained from the parametrization of the
experimental data by Donnachie and Landshoff\cite{DL}, and
quantity $R=\sigma_L(x,Q^2)/\sigma_T(x,Q^2)$
comes from the global fit of the experimental data by Whitlow
{\it et al}\cite{WHIT}.  The result shows that $I^{nonres}(Q^2)$
gives an dominant contribution, while $I^{res}(Q^2)$ only accounts
for 15 to 20 percent of $I(Q^2)$ in the region
$1.0\le Q^2\le 2.5$ GeV$^2$.  This is where we believe
that one could obtain a reliable information of the high
twist effects.  If we write the $Q^2$ dependence of $I(Q^2)$ above
$1$ GeV$^2$ as
\begin{equation}\label{9}
I_p(Q^2)=a+\frac b{Q^2},
\end{equation}
our result in $1\le Q^2\le 2.5$ GeV$^2$ region gives
\begin{equation}\label{10}
a\approx 0.139 \sim 0.144
\end{equation}
and
\begin{equation}\label{11}
b\approx -0.03 \sim -0.04 \mbox{GeV}^2.
\end{equation}
In particular, the parameter $b$ is consistent with
recent analysis of the SMC and EMC data
 by Close and Roberts\cite{CR93} in which they conclude
\begin{equation}\label{29}
b=-0.161\pm 0.530 \mbox{GeV}^2
\end{equation}
 as well as
the QCD sum rule prediction by Balitsky {\it et al}\cite{BAL}
used in the analysis by Ellis and Karliner\cite{EK93}.

\subsection*{\bf 4. Conclusion}
We have presented a framework to study the $Q^2$ dependence of the
spin structure function in the resonance region.  At the real
photon and large $Q^2$ limit, the spin dependent sum rule
$I_p(Q^2)$ defined in Eq. \ref{25} is related the DHG sum rule
and the sum rules in the deep inelastic scattering by
\begin{equation}\label{31}
I_p(Q^2)=\left \{ \begin{array}{c} -\frac {\nu_{th}}{4M}
\kappa^2 \quad Q^2=0 \\ \Gamma \qquad Q^2\to \infty \end{array}
\right .
\end{equation}
which also equals to the integral $\int_0^1 g_1(x,Q^2)dx$
in these limits.  We show that
the data from the resonance region alone is not enough to
determine the integral of the spin structure function,
even the DHG sum rule in the real photon limit.  However,
the spin structure function in the resonance region does
provides important insights into the transition
of the spin structure function
as the $Q^2$ increases.  The DHG sum rule
is a threshold effect. It decreases very quickly as $Q^2$
increases due to its strong connection with the process
$\gamma_v N\to P_{33}(1232)$, which is proportional to
$Q^{-8}$  at large $Q^2$\cite{ZPLI93}, thus it is not
the hight twist effect that
generates the leading $1/Q^2$ corrections
to the spin structure functions in the deep inelastic
scattering region.

Further theoretical and experimental studies of
the quantity $\sigma_{TS}(\nu,Q^2)$ is needed,
it will not only provide information on the spin structure function
$g_2(x,Q^2)$ in the small $Q^2$ region,  but also
allow more precise estimation on the $Q^2$ dependence
of the spin structure function $g_1(x,Q^2)$.  This
investigation is in progress.  Further information on
the asymmetry $A_1(x,Q^2)$ in the small $x$ region around
$Q^2=0$ is desirable to give a precise estimate of
the DHG sum rule, it may also tell us how the transition
from negative to positive $A_1(x,Q^2)$ happens quantitatively
in the small $x$ region.

We show that there is a region that both contributions from resonance
and the nonresonance regions are important, and the effects of the
DHG sum rule is small.  Therefore, one could obtain the leading
high twist effects reliably. Indeed, our result of the $Q^2$ dependence
of $I(Q^2)$ is quite consistent with the recent theoretical analysis.
The future experiments at CEBAF\cite{BEXP} will certainly provide
us more information in this area.

Discussions with V. Burkert are gratefully acknowledged.
This work is supported by the NSF grant PHY-9023586.

\newpage
\section*{Figure Caption}
\begin{enumerate}
\item The shade area is the region of quantity $x$ that can be assessable
to the resonance region.
\item The $Q^2$ evolution of the spin asymmetry function $A_1(x,Q^2)$
in the resonance region; the dot-dashed, short-dashed and solid lines
correspond to $A_1(x,Q^2)$ \cite{AO} at $Q^2=0$, $0.5$ and $1.5$ GeV$^2$
respectively.
The dash-line is the asymmetry $A_1(x)$ in the scaling limit\cite{CLOSE93},
and the data are from Ref. \cite{BAUM}, which the circle(triangle)
represents $A_1(x,Q^2)$ at $Q^2=0.5$($1.5$) GeV$^2$.
\item The function $I(x_{min})$ as a function of $x_{min}$ at
$Q^2=0$. The dashed lines are extensions of $I(x_{min})$ to
$I(x_{min}=0)$ see text.
\item The $Q^2$ dependence of the integral $I^p(Q^2)$
the dashed line corresponds to the contribution from the resonance
region only, and the solid line is the total result, see text.
\end{enumerate}
\end{document}